\documentclass[aps,twocolumn,nofootinbib]{revtex4}
\usepackage{bm,epsf}
\newcommand{\kB}{k_{\rm B}}
\newcommand{\partialr}{\frac{\partial}{\partial\bm{r}}}

\newcommand{\ave}[1]{\left\langle#1\right\rangle}

\begin{document}
\bibliographystyle{apsrev}

\title{Entropic 13 Moment Equations from Boltzmann's Kinetic Equation}

\author{Hans Christian \"Ottinger}
\email[]{hco@mat.ethz.ch}
\homepage[]{http://www.polyphys.mat.ethz.ch/}
\affiliation{ETH Z\"urich, Department of Materials, Polymer Physics, HCI H 543,
CH-8093 Z\"urich, Switzerland}

\date{\today}

\begin{abstract}
We use guiding principles from nonequilibrium thermodynamics to develop an admissible set of $13$ moment equations for rarefied gas flows. The main benefits of our thermodynamic approach are an explicit entropy expression fulfilling an $H$ theorem and a sound Hamiltonian formulation of the reversible free flight transport. To calculate the entropy and to find explicit closure approximations, we propose a simple set of approximate $13$ parameter solutions to Boltzmann's kinetic equation. We discuss how standard hydrodynamics is recovered as a limiting case.
\end{abstract}



\maketitle


\emph{Introduction.}---Rarefied gas flows play an important role in a variety of applications ranging from microfluidics to aerodynamics of satellites and space stations in the outer atmosphere. For these applications, the Navier-Stokes-Fourier equations of hydrodynamics are insufficient and the use of Boltzmann's kinetic equation is computationally too time consuming. Moment equations hence offer an attractive intermediate level of description for rarefied gas flow.


The idea of moment equations goes back to Grad's pioneering work \cite{Grad49,Grad58encyc} on solving Boltzmann's kinetic equation. For a description of both momentum and heat flux, one usually employs $13$ moments. In recent years, a practical, stable, and accurate set of regularized $13$ moment equations has been developed by combining the ideas of the Chapman-Enskog and Grad methods (see, for example, the textbook \cite{Struchtrup}, the letter \cite{StruchtrupTorr07}, and references therein). However, several fundamental issues are still open, most prominently, the existence of an entropy beyond the linear case and the hyperbolicity of the description of free flight transport \cite{Torrilhon10}. Closure based on maximum entropy distributions has been suggested to solve these remaining problems \cite{Levermore96}, but this idea seems to be limited to the $10$ moment case, that is, to the absence of heat flow \cite{Junk98,hcobet}.

The purpose of this work is to develop a fully nonlinear set of $13$ moment equations with the help of nonequilibrium thermodynamics \cite{BerisEdwards,hcobet,hco99,hco100}. We here use two guiding principles: (i) the \emph{existence of an entropy} conserved by reversible processes, and (ii) the possibility of a \emph{Hamiltonian formulation} of reversible dynamics. For our final equations, we obtain an entropy satisfying an $H$ theorem and a proper mechanical description of the reversible evolution associated with free flights.


\emph{Variables.}---We start from Boltzmann's kinetic equation for the single-particle distribution function $f=f(\bm{r},\bm{p})$, that is, the probability density for finding a particle with momentum $\bm{p}$ at the position $\bm{r}$. The normalization is such that $n(\bm{r}) = \int f(\bm{r},\bm{p})d^3p$ is the particle number density. Angular brackets $\ave{\ldots}$ denote position dependent momentum space averages performed with the normalized probability density $f/n$.

We here choose the mass density $\rho = n m$ (where $m$ is the particle mass), the momentum density $\bm{M} = n \ave{\bm{p}}$, and the symmetric second moment tensor $\bm{\pi} = \ave{\bm{p}\bm{p}} - \ave{\bm{p}} \ave{\bm{p}}$ as independent fields. From the third moment tensor of rank three, $\bm{Q} = \ave{ (\bm{p} - \ave{\bm{p}}) (\bm{p} - \ave{\bm{p}}) (\bm{p} - \ave{\bm{p}})}$, we construct the vector $\bm{q} = \bm{Q} : \bm{\pi}^{-1}$ to arrive at our complete set of $13$ fields $(\rho, \bm{M}, \bm{\pi}, \bm{q})$ for which we seek evolution equations. Note that the third-rank tensor $\bm{Q}$ is double contracted with $\bm{\pi}^{-1}$ rather than the unit tensor $\bm{1}$. Whereas the latter option would be motivated by the form of the heat flux vector, our choice of $\bm{q}$ makes the Hamiltonian formulation of reversible dynamics more transparent. The use of general vectors and tensors (rather than Cartesian ones), with the possibility of distinguishing between contravariant and covariant objects, is known to be essential for a deep analysis of deformation and flow \cite{Lodge}. Just like $\bm{M}$, the variable $\bm{q}$ is introduced as a covariant general vector, and $\bm{\pi}$ is a covariant tensor.


\emph{Energy and entropy.}---For our two guiding principles from nonequilibrium thermodynamics, energy and entropy play important roles. In terms of our independent system variables, the total kinetic energy of all noninteracting particles in the volume $V$ occupied by our rarefied gas is given by
\begin{equation}\label{energy}
    E = \int_V \left( \frac{\bm{M}^2}{2\rho} + \frac{\rho}{2m^2}
    {\rm tr}\,\bm{\pi} \right) d^3r .
\end{equation}
For the nonequilibrium entropy of the ideal gas on the $13$ moment level, we assume the general form
\begin{equation}\label{entropy}
    S = \frac{\kB}{m} \int\limits_V \! \left\{ \frac{1}{2}\ln \left[
    \left( \frac{2\pi}{h^2} \right)^{\!\! 3} \!
    \frac{m^2}{\rho^2} \det\bm{\pi} \right] + \frac{5}{2}
    + \bar{S}(\varphi) \right\} \rho \, d^3r ,
\end{equation}
where $\kB$ and $h$ are Boltzmann's  and Planck's constants, respectively, and $\varphi = \bm{q} \cdot \bm{\pi}^{-1} \cdot \bm{q}$ is a dimensionless scalar. We have found this form of the entropy for several classes of skewed trial solutions to Boltzmann's kinetic equation. Before we justify this \emph{ansatz} and calculate the skewness contribution $\bar{S}(\varphi)$ to the entropy per particle in units of $\kB$ for a particular class of trial functions, we give the derivatives of the entropy with respect to $\bm{\pi}$ and $\bm{q}$:
\begin{equation}\label{entropgrad1}
    \frac{\delta S}{\delta \bm{\pi}} =  n \kB \cdot \left[ \frac{1}{2} \bm{\pi}^{-1}
    + b(\varphi) \, \bm{\pi}^{-1} \cdot \bm{q}\bm{q} \cdot \bm{\pi}^{-1} \right] ,
\end{equation}
and
\begin{equation}\label{entropgrad2}
    \frac{\delta S}{\delta \bm{q}} = - 2 n \kB b(\varphi) \, \bm{\pi}^{-1}
    \cdot \bm{q} ,
\end{equation}
with $b(\varphi) = - d \bar{S}(\varphi) / d \varphi \geq 0$.


\emph{A class of 13 parameter functions.}---We next derive the entropy (\ref{entropy}) for a natural class of approximate solutions to Boltzmann's kinetic equation. Our construction of approximate solutions to Boltzmann's kinetic equation is based on multivariate Gaussian distributions (see, for example, Sec.~2.1.3 of \cite{hcobook}),
\begin{equation}\label{Gaussfdef}
    f_{\bm{\lambda}\bm{\Theta}}(\bm{p}) = \frac{1}{{\cal N}}
    \, \exp \left\{ - \frac{1}{2} (\bm{p} - \bm{\lambda}) \cdot
    \bm{\Theta}^{-1} \cdot (\bm{p} - \bm{\lambda}) \right\} ,
\end{equation}
with a first-moment vector $\bm{\lambda}$, a positive-definite symmetric second-moment tensor $\bm{\Theta}$, and the normalization constant ${\cal N} = (2\pi)^{3/2} (\det\bm{\Theta})^{1/2}$. As a simpler alternative to the Pearson-Type-IV distribution considered by Torrilhon \cite{Torrilhon10}, we here propose to use the following class of skewed single-particle distribution functions,
\begin{eqnarray}
    f(\bm{p}) &=& n \Big[ 1 - \bm{w} \cdot \bm{\Theta}^{-1} \cdot
    (\bm{p} - \bm{\lambda})
    \nonumber \\
    & \times & a \big( (\bm{p} - \bm{\lambda}) \cdot
    \bm{\Theta}^{-1} \cdot (\bm{p} - \bm{\lambda}) \big) \Big] \,
    f_{\bm{\lambda}\bm{\Theta}}(\bm{p}) ,
\label{13momfunctions}
\end{eqnarray}
where a suitably chosen function $a(x)$ regularizes the skewness introduced by $\bm{w}$. A convenient choice of $a(x)$ is discussed below. Note that $f$ is nonnegative if $\vartheta = \bm{w} \cdot \bm{\Theta}^{-1} \cdot \bm{w} \leq [p a(p^2)]^{-2}$ for all $p$; it is hence important that there exists an upper bound for $|p a(p^2)|$ and, therefore, that $a(p^2)$ is nonpolynomial. In three dimensions, the scalar number density $n$, the vector $\bm{\lambda}$, the symmetric tensor $\bm{\Theta}$, and the skewness vector $\bm{w}$ add up to a total of $13$ parameters. These parameters are actually fields depending on the position $\bm{r}$. The single-particle distribution functions (\ref{13momfunctions}) can be thought of as trial solutions to Boltzmann's kinetic equation or as an approximate invariant manifold \cite{GorbanKarlin94a,GorbanKarlin,hco190}.

A nice feature of our trial functions (\ref{13momfunctions}) is that all even moments coincide with those of the Gaussian distribution $f_{\bm{\lambda}\bm{\Theta}}$. The odd moments can be reduced to more complicated Gaussian averages involving $a(x)$. For the lowest three moments, we obtain the closed-form expressions
\begin{equation}\label{mom12}
    \bm{M} = n ( \bm{\lambda} - a_4 \, \bm{w} ) , \qquad
    \bm{\pi} = \bm{\Theta} - a_4^2 \, \bm{w}\bm{w} ,
\end{equation}
\begin{equation}\label{mom3q}
    \bm{q} = \theta(\vartheta) \, \bm{w} , \qquad
    \theta(\vartheta) = \left( 4 a_4 - 2 a_6 -
    \frac{3 a_6 - a_4}{1 - a_4^2 \vartheta} \right) ,
\end{equation}
with the numerical coefficients
\begin{equation}\label{atransform}
    a_j = \frac{1}{\sqrt{2\pi} \, (j-1)!!}
    \int_{-\infty}^\infty p^j \, a(p^2) \, e^{-p^2/2} dp ,
\end{equation}
where $k!!$ is the product of all odd numbers from $1$ to $k$. A convenient criterion for the choice of $a(x)$ is that the numbers $a_j$ can be evaluated in closed form. For $a(x) = \exp(-u x/2)$, with a parameter $u>0$, we have $a_j = (1+u)^{-(j+1)/2}$ and hence $a_4/a_6 = 1+u$. Once the relationship between the scalars $\vartheta = \bm{w} \cdot \bm{\Theta}^{-1} \cdot \bm{w}$ and $\varphi = \bm{q} \cdot \bm{\pi}^{-1} \cdot \bm{q}$ is established,
\begin{equation}\label{phithetarel}
    \varphi = \frac{\vartheta}{1 - a_4^2 \vartheta} \, \theta(\vartheta)^2 ,
\end{equation}
the inversion of Eqs.~(\ref{mom12})--(\ref{mom3q}) to obtain the parameters of the trial solutions in terms of the moments is simple.

\begin{figure}
\centerline{\epsfxsize=6cm \epsffile{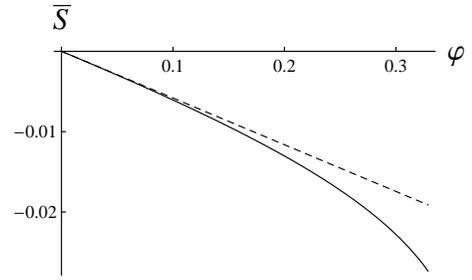}}
\caption[ ]{Skewness contribution $\bar{S}(\varphi)$ to the entropy (continuous line) and linear approximation to it (dashed line).}
\label{fig_entropy}
\end{figure}

By integrating $-f \ln f$ and using the relationship $\det\bm{\pi} = (1 - a_4^2 \vartheta) \det\bm{\Theta}$, we confirm Eq.~(\ref{entropy}) and find the following skewness contribution to the entropy:
\begin{eqnarray}
    \bar{S}(\vartheta) &=& \frac{-1}{\sqrt{2\pi}} \int_0^\infty \bigg\{ \frac{1}{2}
    \left( zpa + \frac{1}{zpa} \right) \ln \frac{1+zpa}{1-zpa}
    \nonumber \\
    && \hspace{-3.5em} + \ln [1-(zpa)^2] - 1 \bigg\} \, p^2 e^{-p^2/2} dp
    - \frac{1}{2} \ln(1 - a_4^2 \vartheta) , \qquad
\label{entropybar}
\end{eqnarray}
where $z=\sqrt{\vartheta}$ and $a=a(p^2)$. For our example $a(x) = \exp(-u x/2)$ with the value $u=2/5$, suggested to reproduce the trial solutions of Grad's approach upon Taylor expansion of $a(x)$, the corresponding entropy contribution $\bar{S}(\varphi)$, as obtained by numerical integration and the relation (\ref{phithetarel}) between $\vartheta$ and $\varphi$, is represented by the continuous line in Fig.~\ref{fig_entropy}. The entropy clearly favors the symmetric distributions characterized by $\varphi=0$. The linear approximation $\bar{S}(\varphi) = - b(0) \, \varphi$ with
\begin{equation}\label{linentropapp}
    b(0) = \frac{1}{50} \left\{ \left[ 1 + \frac{u^2}{1+2u} \right]^{5/2} - 1 \right\}
    \left(1 + \frac{1}{u}\right)^2 ,
\end{equation}
implying a quadratic dependence of $\bar{S}$ on $\bm{q}$, is shown by the dashed line in Fig.~\ref{fig_entropy}. It provides a good approximation over the entire range of allowed values of $\varphi$. In the limit of small $u$, $b(0)$ is $1/20$; for $u=2/5$, we have $b(0) \approx 0.058$.


\emph{Evolution equations.}---With an entropy at hand, we can now use our thermodynamic guiding principles to formulate evolution equations for our $13$ moments. In addition to the usual continuity and momentum balance equations,
\begin{equation}\label{rhoMevol}
    \frac{\partial\rho}{\partial t} = - \partialr \cdot (\bm{v} \rho) , \quad
    \frac{\partial\bm{M}}{\partial t} = - \partialr \cdot
    \left( \bm{v} \bm{M} + \frac{\rho}{m^2} \bm{\pi} \right) , \,\,
\end{equation}
where $\bm{v} = \bm{M}/\rho$ is the velocity field, we have the following evolution equation for the second moment tensor $\bm{\pi}$
\begin{eqnarray}
    \frac{\partial\bm{\pi}}{\partial t} &=& - \bm{v} \cdot \partialr \bm{\pi}
    - \bm{\kappa} \cdot \bm{\pi} - \bm{\pi} \cdot \bm{\kappa}^T
    - \frac{1}{m \rho}\partialr \cdot (\rho \bm{Q})
    \nonumber \\
    &-& \frac{1}{\tau} \left( \bm{\pi}  - \frac{{\rm tr}\bm{\pi}}{3} \, \bm{1} \right) ,
\label{pievol}
\end{eqnarray}
where we have introduced the velocity gradient tensor with components $\kappa_{jk} = \partial v_j/\partial r_k$. The reversible terms in the first line are exact, but we still need a closure approximation for the third moment tensor $\bm{Q}$. From the general tensor structure, we expect
\begin{equation}\label{mom3piq}
    Q_{ijk} =  \bar{Q}_1 ( q_i \pi_{jk} + q_j \pi_{ik} + q_k \pi_{ij} )
    - \bar{Q}_2 \, q_i q_j q_k ,
\end{equation}
with suitable functions $\bar{Q}_1$ and $\bar{Q}_2$ of $\varphi$ (or, in view of Eq.~(\ref{phithetarel}), equivalently of $\vartheta$). For our trial functions (\ref{13momfunctions}), we obtain
\begin{equation}\label{Qbarexpr}
    \bar{Q}_1 = \frac{a_4-a_6}{\theta(\vartheta)} , \qquad
    \bar{Q}_2 = \frac{(3a_6-a_4) \, a_4^2}{\theta(\vartheta)^3} .
\end{equation}
For $\vartheta=\varphi=0$, we find $\bar{Q}_1 = 1/5$, and we suggest that $2/5$ is a realistic choice of $\bar{Q}_2$. Alternatively, $\bar{Q}_1$ and $\bar{Q}_2$ could be obtained by matching simulation results or some established $13$ moment equations. We here do not discuss further restrictions for the most general functional form of $\bar{Q}_1$ and $\bar{Q}_2$ that might result from the Hamiltonian character of reversible dynamics. The relaxation term in Eq.~(\ref{pievol}) has the simplest possible form consistent with energy conservation in the collision process.

We now turn to the evolution of $\bm{q}$. From the perspective of kinetic theory, $\bm{q}$ is a well-defined combination of second and third moments. From a phenomenological perspective, we only know that $\bm{q}$ is a covariant vector. A clear physical meaning is given to $\bm{q}$ through the closure (\ref{mom3piq}) which relates $\bm{q}$ to the heat flux,
\begin{equation}\label{heatfluxexp}
    \bm{j}^{\rm q} = \frac{\rho}{2m^3} \, \bm{Q} : \bm{1} .
\end{equation}
We could, for example, fix the meaning of $\bm{q}$ by choosing $\bar{Q}_1$ to be identically equal to $1/5$. As an alternative, we could make $\bm{q}$ phenomenologically well-defined by choosing the form of the entropy contribution $\bar{S}(\varphi)$, for example, by choosing $b$ to be constant (say, $1/20$). As there is no straightforward phenomenological relationship between third moments and entropy, we might even want to prescribe both $\bar{Q}_1$ and $b$. As an evolution equation for the skewness vector $\bm{q}$, we now propose
\begin{eqnarray}
    \frac{\partial\bm{q}}{\partial t} &=&
    - \bm{v} \cdot \partialr \bm{q} - \bm{\kappa} \cdot \bm{q}
    - \frac{1}{2m \rho} \bm{q} \cdot \bm{\pi}^{-1} \cdot
    \left[ \partialr \cdot (\rho \bm{Q}) \right]
    \nonumber \\
    &-& \frac{1}{4m \rho \varphi b} \, \bm{q} \, \bm{\pi}^{-1} :
    \left[ \partialr \cdot (\rho \bm{Q}) \right] \label{wevol} \\
    &-& \frac{F}{m} \left( \bm{1} - \frac{\bm{q}\bm{q}}{\varphi}
    \cdot \bm{\pi}^{-1} \right) \cdot \partialr \,
    {\rm tr}\bm{\pi}  - \frac{1}{2\tau} \bm{D} \cdot \bm{q} .
    \nonumber
\end{eqnarray}
All reversible terms in the first two lines of Eq.~(\ref{wevol}) are dictated by our first thermodynamic guiding principle, that is, the conservation of the entropy (\ref{entropy}) by reversible dynamics. The exact evolution equation for $\bm{q}$ involves divergences of the third and fourth moment tensors. The terms involving the divergence of $\rho \bm{Q}$ in Eq.~(\ref{wevol}) have exactly the same tensor structure as in the exact evolution equation. The reversible term involving the scalar fitting parameter $F$ represents the effect of fourth moments. As this effect occurs only in the evolution of $\bm{q}$ but not of $\bm{\pi}$, it is highly restricted by our second guiding principle, the Hamiltonian character of reversible dynamics. The proposed term is the only one we could come up with in view of our two guiding principles. Whereas it does not match the full structure of the terms resulting from fourth-moment closure, matching the hydrodynamic limit suggests $F=5/3$. Finally, the relaxation term in Eq.~(\ref{wevol}) involves a tensor $\bm{D}$.

From the evolution equations (\ref{rhoMevol}), (\ref{pievol}), and (\ref{wevol}) we obtain the entropy balance equation
\begin{equation}\label{sevol}
    \frac{\partial s}{\partial t} = - \partialr \cdot (\bm{v} s)
    + \frac{n \kB}{6 \tau} ( {\rm tr}\bm{\pi} \, {\rm tr}\bm{\pi}^{-1} - 9 )
    + \frac{n \kB b}{\tau} \bm{q} \cdot \bm{\pi}^{-1} \cdot \bar{\bm{D}} \cdot \bm{q} ,
\end{equation}
where $\bar{\bm{D}} = \bm{D} - \bm{1} + ({\rm tr}\bm{\pi}/3) \bm{\pi}^{-1}$ and $s$ is the entropy density implied by Eq.~(\ref{entropy}). For positive definite $\bm{\pi}^{-1} \cdot \bar{\bm{D}}$, we have thus established an $H$ theorem for the thermodynamically admissible $13$ moment equations. From now on, we simply assume an isotropic tensor $\bar{\bm{D}} = \bar{D} \bm{1}$.


\emph{Hydrodynamic limit.}---Within the Chapman-Enskog procedure for obtaining hydrodynamic equations in the limit of short relaxation times $\tau$, Eq.~(\ref{pievol}) leads to
\begin{equation}\label{pievolCE}
    \bm{\pi}  - \frac{{\rm tr}\bm{\pi}}{3} \, \bm{1} = - m \kB T \,
    \tau ( \bm{\kappa} + \bm{\kappa}^T ) ,
\end{equation}
where $T$ is the local equilibrium temperature implied by the trace of $\bm{\pi}$. Equation (\ref{pievolCE}) represents Newton's expression for the stress tensor. For the discussion of heat flow, it is useful to look at the evolution equation for the scalar $\varphi$ obtained by combining Eqs.~(\ref{pievol}) and (\ref{wevol}),
\begin{equation}\label{phievol}
    \frac{\partial\varphi}{\partial t} =
    - \bm{v} \cdot \frac{\partial\varphi}{\partial\bm{r}}
    - \frac{1}{2m \rho b} \, \bm{\pi}^{-1} :
    \left[ \partialr \cdot (\rho \bm{Q}) \right]
    - \frac{\bar{D}}{\tau} \, \varphi .
\end{equation}
The Chapman-Enskog procedure then leads to
\begin{equation}\label{phievolCE}
    \varphi = - \frac{\tau}{n \kB T b \bar{D}} \partialr \cdot \bm{j}^{\rm q} .
\end{equation}
After introducing this limiting form of $\varphi$ into the last term of the entropy balance (\ref{sevol}), the proper combination of the thermal entropy flux and production arises.

When Eq.~(\ref{phievolCE}) is used in the further analysis of Eq.~(\ref{wevol}), one can recognize that the term with prefactor $F$ must become small. This observation implies that the temperature gradient becomes equal to its projection along $\bm{q}$, that is, $\bm{q}$ must be proportional to the temperature gradient. In other words, the direction of $\bm{q}$ is found by suppressing components perpendicular to the temperature gradient.


\emph{Summary and outlook.}---The evolution equations (\ref{rhoMevol}), (\ref{pievol}), and (\ref{wevol}) developed in this letter may be considered as thermodynamically closed $13$ moment equations for rarefied gas flow. They come with the nondecreasing entropy (\ref{entropy}) and a physically appealing Hamiltonian formulation of reversible dynamics. Moreover, standard hydrodynamics is recovered as a limiting case. As thermodynamics is designed exactly for that purpose, we expect well-behaved solutions in all physically meaningful applications of these equations. However, a detailed study of solutions is beyond the scope of this fundamental development of thermodynamically admissible equations.

From the perspective of thermodynamic modeling of skewness on the $13$ moment level, the numbers or functions $\bar{Q}_1$, $\bar{Q}_2$, and $F$ arising from the third and fourth moment closure and the entropy contribution $\bar{S}(\varphi)$ [or, equivalently, its derivative $b(\varphi)$ occurring in the evolution equation (\ref{wevol})] may be considered as independent inputs. From the perspective of kinetic theory, these inputs (with a grain of salt in the case of $F$) are related to the function $a(x)$ in the trial functions (\ref{13momfunctions}), namely by Eqs.~(\ref{atransform}) and (\ref{entropybar}). As long as there is no profoundly justified choice for $a(x)$, however, one should feel free to fit all thermodynamic parameters or to use a simple set of parameters, for example $\bar{Q}_1 = 1/5$, $\bar{Q}_2 = 2/5$, $F=5/3$, and $b=1/20$. The dissipative parameters $\tau$ and $\bar{D}$ can be obtained by matching our evolution equations with kinetic theory or experimental results. For Maxwell molecules, we expect $\bar{D} = 4/3$ \cite{Grad58encyc,Struchtrup,StruchtrupTorr07}.

For most applications, the $13$ moment equations need to be supplemented by boundary conditions. In the work \cite{StruchtrupTorr07}, the usefulness of thermodynamics in providing meaningful boundary conditions has been recognized. It may hence be worthwhile to point out that the GENERIC framework, employed here to obtain the bulk equations, has recently been supplemented by boundary thermodynamics \cite{hco162,hco188}.

The treatment of rarefied gases on the level of $13$ moments, which is intermediate between kinetic theory and hydrodynamics, is a classical example of multiscale modeling. In addition to the thermodynamic exploration of this intermediate level, we have connected its parameters to kinetic theory. Our remarkably simple equations (\ref{rhoMevol}), (\ref{pievol}), and (\ref{wevol}) show that thermodynamic principles, although strictly speaking never indispensable, offer enormously useful guidance.

I gratefully acknowledge inspiring discussions with Henning Struchtrup, Manuel Torrilhon, and Martin Kr\"oger.



\begin{thebibliography}{18}
\expandafter\ifx\csname natexlab\endcsname\relax\def\natexlab#1{#1}\fi
\expandafter\ifx\csname bibnamefont\endcsname\relax
  \def\bibnamefont#1{#1}\fi
\expandafter\ifx\csname bibfnamefont\endcsname\relax
  \def\bibfnamefont#1{#1}\fi
\expandafter\ifx\csname citenamefont\endcsname\relax
  \def\citenamefont#1{#1}\fi
\expandafter\ifx\csname url\endcsname\relax
  \def\url#1{\texttt{#1}}\fi
\expandafter\ifx\csname urlprefix\endcsname\relax\def\urlprefix{URL }\fi
\providecommand{\bibinfo}[2]{#2}
\providecommand{\eprint}[2][]{\url{#2}}

\bibitem[{\citenamefont{Grad}(1949)}]{Grad49}
\bibinfo{author}{\bibfnamefont{H.}~\bibnamefont{Grad}},
  \bibinfo{journal}{Commun.\ Pure Appl.\ Math.} \textbf{\bibinfo{volume}{2}},
  \bibinfo{pages}{331} (\bibinfo{year}{1949}).

\bibitem[{\citenamefont{Grad}(1958)}]{Grad58encyc}
\bibinfo{author}{\bibfnamefont{H.}~\bibnamefont{Grad}}, in
  \emph{\bibinfo{booktitle}{Thermodynamics of Gases}}, edited by
  \bibinfo{editor}{\bibfnamefont{S.}~\bibnamefont{Fl{\"u}gge}}
  (\bibinfo{publisher}{Springer}, \bibinfo{address}{Berlin},
  \bibinfo{year}{1958}), no. \bibinfo{number}{Volume XII} in
  \bibinfo{series}{Encyclopedia of Physics}, pp. \bibinfo{pages}{205--294}.

\bibitem[{\citenamefont{Struchtrup}(2005)}]{Struchtrup}
\bibinfo{author}{\bibfnamefont{H.}~\bibnamefont{Struchtrup}},
  \emph{\bibinfo{title}{Macroscopic Transport Equations for Rarefied Gas
  Flows}} (\bibinfo{publisher}{Springer}, \bibinfo{address}{Berlin},
  \bibinfo{year}{2005}).

\bibitem[{\citenamefont{Struchtrup and Torrilhon}(2007)}]{StruchtrupTorr07}
\bibinfo{author}{\bibfnamefont{H.}~\bibnamefont{Struchtrup}} \bibnamefont{and}
  \bibinfo{author}{\bibfnamefont{M.}~\bibnamefont{Torrilhon}},
  \bibinfo{journal}{Phys.\ Rev.\ Lett.} \textbf{\bibinfo{volume}{99}},
  \bibinfo{pages}{014502} (\bibinfo{year}{2007}).

\bibitem[{\citenamefont{Torrilhon}(2010)}]{Torrilhon10}
\bibinfo{author}{\bibfnamefont{M.}~\bibnamefont{Torrilhon}},
  \bibinfo{journal}{Commun.\ Comput.\ Phys.} \textbf{\bibinfo{volume}{7}},
  \bibinfo{pages}{639} (\bibinfo{year}{2010}).

\bibitem[{\citenamefont{Levermore}(1996)}]{Levermore96}
\bibinfo{author}{\bibfnamefont{C.~D.} \bibnamefont{Levermore}},
  \bibinfo{journal}{J.~Stat.\ Phys.} \textbf{\bibinfo{volume}{83}},
  \bibinfo{pages}{1021} (\bibinfo{year}{1996}).

\bibitem[{\citenamefont{Junk}(1998)}]{Junk98}
\bibinfo{author}{\bibfnamefont{M.}~\bibnamefont{Junk}},
  \bibinfo{journal}{J.~Stat.\ Phys.} \textbf{\bibinfo{volume}{93}},
  \bibinfo{pages}{1143} (\bibinfo{year}{1998}).

\bibitem[{\citenamefont{{\"O}ttinger}(2005)}]{hcobet}
\bibinfo{author}{\bibfnamefont{H.~C.} \bibnamefont{{\"O}ttinger}},
  \emph{\bibinfo{title}{Beyond Equilibrium Thermodynamics}}
  (\bibinfo{publisher}{Wiley}, \bibinfo{address}{Hoboken},
  \bibinfo{year}{2005}).

\bibitem[{\citenamefont{Beris and Edwards}(1994)}]{BerisEdwards}
\bibinfo{author}{\bibfnamefont{A.~N.} \bibnamefont{Beris}} \bibnamefont{and}
  \bibinfo{author}{\bibfnamefont{B.~J.} \bibnamefont{Edwards}},
  \emph{\bibinfo{title}{The Thermodynamics of Flowing Systems}}
  (\bibinfo{publisher}{Oxford University Press}, \bibinfo{address}{New York},
  \bibinfo{year}{1994}).

\bibitem[{\citenamefont{Grmela and {\"O}ttinger}(1997)}]{hco99}
\bibinfo{author}{\bibfnamefont{M.}~\bibnamefont{Grmela}} \bibnamefont{and}
  \bibinfo{author}{\bibfnamefont{H.~C.} \bibnamefont{{\"O}ttinger}},
  \bibinfo{journal}{Phys.\ Rev.\ E} \textbf{\bibinfo{volume}{56}},
  \bibinfo{pages}{6620} (\bibinfo{year}{1997}).

\bibitem[{\citenamefont{{\"O}ttinger and Grmela}(1997)}]{hco100}
\bibinfo{author}{\bibfnamefont{H.~C.} \bibnamefont{{\"O}ttinger}}
  \bibnamefont{and} \bibinfo{author}{\bibfnamefont{M.}~\bibnamefont{Grmela}},
  \bibinfo{journal}{Phys.\ Rev.\ E} \textbf{\bibinfo{volume}{56}},
  \bibinfo{pages}{6633} (\bibinfo{year}{1997}).

\bibitem[{\citenamefont{Lodge}(1974)}]{Lodge}
\bibinfo{author}{\bibfnamefont{A.~S.} \bibnamefont{Lodge}},
  \emph{\bibinfo{title}{Body Tensor Fields in Continuum Mechanics}}
  (\bibinfo{publisher}{Academic Press}, \bibinfo{address}{New York},
  \bibinfo{year}{1974}).

\bibitem[{\citenamefont{{\"O}ttinger}(1996)}]{hcobook}
\bibinfo{author}{\bibfnamefont{H.~C.} \bibnamefont{{\"O}ttinger}},
  \emph{\bibinfo{title}{Stochastic Processes in Polymeric Fluids: Tools and
  Examples for Developing Simulation Algorithms}}
  (\bibinfo{publisher}{Springer}, \bibinfo{address}{Berlin},
  \bibinfo{year}{1996}).

\bibitem[{\citenamefont{Gorban and Karlin}(1994)}]{GorbanKarlin94a}
\bibinfo{author}{\bibfnamefont{A.~N.} \bibnamefont{Gorban}} \bibnamefont{and}
  \bibinfo{author}{\bibfnamefont{I.~V.} \bibnamefont{Karlin}},
  \bibinfo{journal}{Physica A} \textbf{\bibinfo{volume}{206}},
  \bibinfo{pages}{401} (\bibinfo{year}{1994}).

\bibitem[{\citenamefont{Gorban and Karlin}(2005)}]{GorbanKarlin}
\bibinfo{author}{\bibfnamefont{A.~N.} \bibnamefont{Gorban}} \bibnamefont{and}
  \bibinfo{author}{\bibfnamefont{I.~V.} \bibnamefont{Karlin}},
  \emph{\bibinfo{title}{Invariant Manifolds for Physical and Chemical
  Kinetics}}, no. \bibinfo{number}{660} in \bibinfo{series}{Lecture Notes in
  Physics} (\bibinfo{publisher}{Springer}, \bibinfo{address}{Berlin},
  \bibinfo{year}{2005}).

\bibitem[{\citenamefont{Colangeli et~al.}(2009)\citenamefont{Colangeli,
  Kr{\"o}ger, and {\"O}ttinger}}]{hco190}
\bibinfo{author}{\bibfnamefont{M.}~\bibnamefont{Colangeli}},
  \bibinfo{author}{\bibfnamefont{M.}~\bibnamefont{Kr{\"o}ger}},
  \bibnamefont{and} \bibinfo{author}{\bibfnamefont{H.~C.}
  \bibnamefont{{\"O}ttinger}}, \bibinfo{journal}{Phys.\ Rev.\ E}
  \textbf{\bibinfo{volume}{80}}, \bibinfo{pages}{051202}
  (\bibinfo{year}{2009}).

\bibitem[{\citenamefont{{\"O}ttinger}(2006)}]{hco162}
\bibinfo{author}{\bibfnamefont{H.~C.} \bibnamefont{{\"O}ttinger}},
  \bibinfo{journal}{Phys.\ Rev.\ E} \textbf{\bibinfo{volume}{73}},
  \bibinfo{pages}{036126} (\bibinfo{year}{2006}).

\bibitem[{\citenamefont{{\"O}ttinger et~al.}(2009)\citenamefont{{\"O}ttinger,
  Bedeaux, and Venerus}}]{hco188}
\bibinfo{author}{\bibfnamefont{H.~C.} \bibnamefont{{\"O}ttinger}},
  \bibinfo{author}{\bibfnamefont{D.}~\bibnamefont{Bedeaux}}, \bibnamefont{and}
  \bibinfo{author}{\bibfnamefont{D.~C.} \bibnamefont{Venerus}},
  \bibinfo{journal}{Phys.\ Rev.\ E} \textbf{\bibinfo{volume}{80}},
  \bibinfo{pages}{021606} (\bibinfo{year}{2009}).

\end{thebibliography}

\end{document}